\documentclass[superscriptaddress,prl,showpacs,twocolumn]{revtex4}%

\usepackage{amsfonts}
\usepackage{amssymb}
\usepackage{amsmath}
\usepackage{graphicx}
\usepackage{epsfig}%

\begin{document}
\title{Observable Topological Effects of M\"{o}bius Molecular Devices}
\author{Nan Zhao}
\affiliation{Institute of Theoretical Physics, Chinese Academy of
Sciences, Beijing, 100190, China} \affiliation{Department of
Physics, Tsinghua University, Beijing 100084, China}
\author{H. Dong}
\affiliation{Institute of Theoretical Physics, Chinese Academy of
Sciences, Beijing, 100190, China}
\author{Shuo Yang}
\affiliation{Institute of Theoretical Physics, Chinese Academy of
Sciences, Beijing, 100190, China}
\author{C. P. Sun}
\affiliation{Institute of Theoretical Physics, Chinese Academy of
Sciences, Beijing, 100190, China}

\begin{abstract}
We study the topological properties of quantum states for the
spinless particle hopping in a M\"{o}bius ladder. This system can
be regarded as  a molecular device possibly engineered from the
aromatic M\"{o}bius annulenes, which enjoys a pseudo-spin orbital
interaction described by a non-Abelian gauge structure. It results
from the nontrivial topology of configuration space, and results
in various observable effects, such as optical spectral splitting.
The transmission spectrum through the M\"{o}bius molecular device
is calculated to demonstrate a topological effect as a destructive
interferences in the conduction band. The induced interaction also
leads to an entanglement between the transverse and longitudinal
modes for any locally factorized state.

\end{abstract}

\maketitle

\textit{Introduction}.- With various potential applications, the
molecular based devices have motivated extensive experimental and
theoretical investigations (see Ref.\cite{bookMoleculardevice}).
Similar to the semiconductor based artificial nanostructures, the
engineered molecular architectures also displays different novel
quantum effects. The concept of M\"{o}bius aromaticity was firstly
proposed in 1964\cite{Heilbronner}, but various kinds of M\"{o}bius
molecule were not designed or claimed to be
synthesized\cite{ChemRev,NatureExperiment} until recent years. With
these significant developments of engineering topologically
nontrivial chemical structures, the topological properties of
quantum systems become more and more important. Thus, it is natural
to consider the various quantum effects induced by the nontrivial
topological configurations\cite{PapersOnMobius1,PapersOnMobius2,
PapersOnMobius3,PapersOnMobius4,PapersOnMobius5,PapersOnMobius6},
such as the M\"{o}bius molecule.

For the topological effects in quantum mechanics, it is a
fundamental principle (assumption) that the wave function must be
single-valued in a topologically trivial configuration. In this
sense, various topological effects such as Aharonov-Bohm (AB) effect
can be rationally explained without the introduction of any extra
assumption in quantum mechanics\cite{CNYang}. In some configuration
with nontrivial topological structure, though the motion of particle
requires some complex boundary conditions, such topological
non-trivialness of the configuration space can be canceled by
introducing a singular gauge field. A typical example is the
phenomena of persistent currents in a mesoscopic or a
superconducting ring threaded by a magnetic flux. Here, the $U(1)$
gauge field can be introduced to cancel the seemingly multi-value of
boundary condition with a non-integrable AB phase factor. Another
illustration is the fractional statistics of anyons, which describes
the effective quantum excitations of the hardcore particles confined
strictly in two dimension with the braiding homotopy of the
many-body configuration \cite{WelcekYSWu}.

In this Letter, we show a topology induced quantum effect with a
non-Abelian gauge structure, which emerges from the twisted
boundary condition in a M\"{o}bius ladder. The twisted boundary
condition results in a local interaction between transverse and
longitudinal modes. And, in the continuous limit, the spinless
particle moving in the M\"{o}bius ladder is mapped to a
pseudo-spin coupled to the orbit corresponding to the longitudinal
mode. When we apply a transverse field, the pseudo-spin seems to
be confined in a one-dimensional ring and subject to a
texture-like effective magnetic field, together with an effective
magnetic flux threading the ring(see Fig.1). Different from the
existing setups of mesoscopic ring for persistent
currents\cite{DLoss}, the effective magnetic flux in our molecular
device depends on the pseudo-spin state, namely, there exists a
non-Abelian gauge field induced by the M\"{o}bius topology.

\begin{figure}[pb]
\includegraphics[bb=29 60 233 772, angle=-90, width=7.5 cm,
clip]{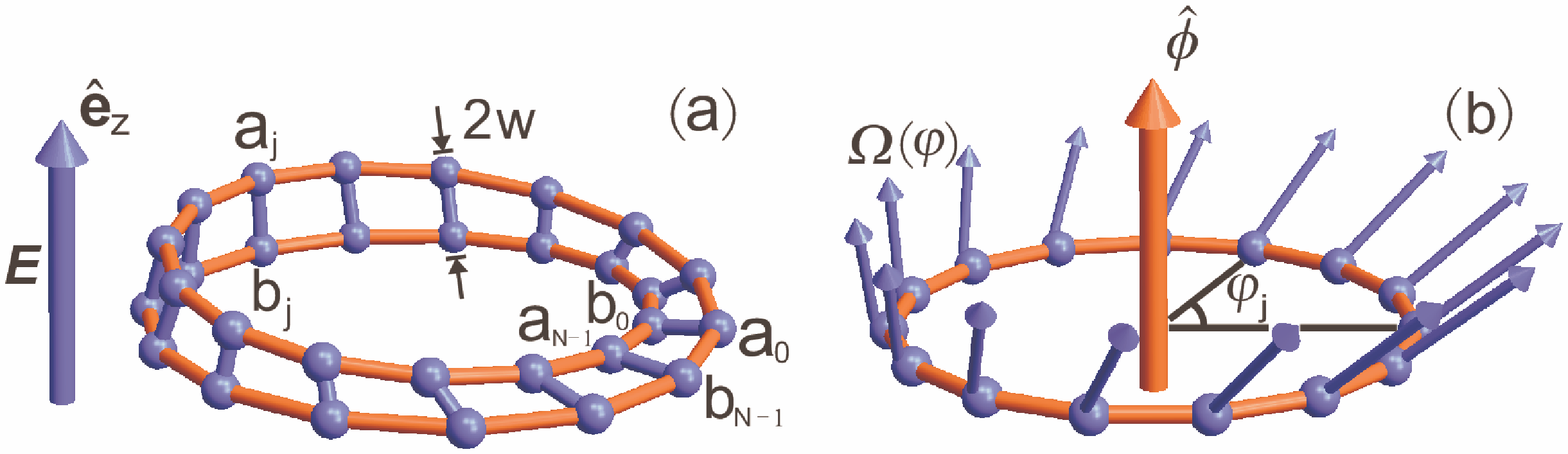} \caption{(a) Schematic illustration
of M\"{o}bius ladder. (b) An electron hopping in the M\"{o}bius
ladder can be mapped to a pseudo-spin moving in a one-dimension
ring, with a spin-dependent flux $\hat{\phi}$ and a texture-like
inhomogeneous magnetic
field $\boldsymbol{\Omega}(\varphi)$.}%
\label{Fig1}%
\end{figure}

Compared with the topologically trivial configurations, i.e. an
ordinary tight binding ladder, the quantum state of the M\"{o}bius
molecule is strongly affected by its configuration topology. Here,
we predict three quantum interference effects, which are
distinguished from the ordinary cases obviously: (i) the Stark
effect in the electric field cause more optical spectral splitting,
due to the effective Zeeman effect by the induce non-Abelian gauge
field; (ii) the transmission spectrum through the M\"{o}bius
molecule is significantly modified due to the destructive
interference caused by the non-Abelian flux; (iii) from the view of
quantum information, the entanglement is emerged from the locally
factorized state in the M\"{o}bius molecule.

\textit{Induced non-Abelian gauge structure}.- Let us consider the
hopping of an electron on the tight-binding lattice with a
M\"{o}bius ladder configuration illustrated in Fig.\ref{Fig1}(a).
Since there is no spin flip, the electron is regarded as a spinless
particle. There are $2N$ lattice sites, located at the two edges of
the ladder, whose coordinates are $\mathbf{r}_{j\pm}=(
\cos\varphi_{j}\left(R\pm w\sin[\varphi_{j}/2]\right),
\sin\varphi_{j}\left(R\pm w\sin[\varphi_{j}/2]\right),\\ \pm
w\cos[\varphi_{j}/2])$, with half-width $w$ and ``radius'' $R$, and
$\varphi_{j}=2\pi j/N$ being the polar coordinate. We introduce
operators $a_{j}^{\dagger}$ ($a_{j}$) and $b_{j}^{\dagger}$
($b_{j}$) to denote the creating (annihilating) a particle on $j$-th
site of each edges respectively. The rungs represent the coupling
between $a$-chain and $b $-chain. We assume the hopping strength
along the ladder is homogeneous. Then the Hamiltonian reads
\begin{equation}
H=\sum_{j=0}^{N-1}\mathbf{A}_{j}^{\dagger}\mathbf{M}_{j}\mathbf{A}_{j}-\xi
\sum_{j=0}^{N-1}\left(\mathbf{A}_{j}^{\dagger}\mathbf{A}_{j+1}+\text{h.c.}\right),
\end{equation}
where $\mathbf{A}_{j}=(a_{j},b_{j})^{\text{T}}$, and the matrix
$\mathbf{M}_{j}=\varepsilon_{j}\sigma_{z}-V_{j}\sigma_{x},$ for
$\sigma_{x,y,z}$ being the Pauli matrices, $2\varepsilon_{j}$
describing the on-site energy difference between $a$- and
$b$-particles, and $V_{j}$ representing their coupling strength.

The M\"{o}bius boundary conditions are $a_{N}=b_{0}$ and
$b_{N}=a_{0}$, or equivalently,
$\mathbf{A}_{N}=\sigma_{x}\mathbf{A}_{0}.$ It is this boundary
condition that results in the interesting topological properties of
the quantum state. In terms of the operator-valued vector
$\mathbf{B}_{j}\equiv
(c_{j\uparrow},c_{j\downarrow})^{\text{T}}=(\exp(-i\varphi
_{j}/2)(a_{j}-b_{j}),a_{j}+b_{j})^{\text{T}}/\sqrt{2}$, which is a
unitary transformation of $\mathbf{A}_{j}$, the Hamiltonian is
rewritten as
\begin{equation}
H=\sum_{j=0}^{N-1}\mathbf{B}_{j}^{\dagger}(\mathbf{\Omega}_{j}\mathbb{\cdot
}\boldsymbol{\sigma})\mathbf{B}_{j}-\xi\sum_{j=0}^{N-1}\left(\mathbf{B}_{j}
^{\dagger}Q\mathbf{B}_{j+1}+\text{h.c.}\right),\label{HaminonAbelian}
\end{equation}
where $\mathbf{\Omega}_{j}\equiv\left(  \varepsilon_{j}\cos(\varphi
_{j}/2),\varepsilon_{j}\sin(\varphi_{j}/2),V_{j}\right) ^{\text{T}}$
is a direction vector, and $Q\equiv\text{diag}[\exp(i\pi/N),1]$. It
should be emphasized that the operator $\mathbf{B}_{j}$ only
requires the ordinary periodic boundary condition, i.e.,
$\mathbf{B}_{N}=\mathbf{B}_{0}.$ So far, we have shown that the
nontrivial M\"{o}bius boundary condition is canceled by the unitary
transformation, accompanied by an induced non-Abelian gauge field
associated with $\mathbf{\Omega}_{j}$ and $Q$. This point will be
seen more clearly in the continuous limit below.

In the continuous limit (i.e. $N\rightarrow\infty$ and
$\varphi_j\rightarrow\varphi\in[0,2\pi]$), the particle hopping on a
M\"{o}bius ladder is described by the two-component Hamiltonian
(\ref{HaminonAbelian}), which can be mapped to the continuous
Hamiltonian
\begin{equation}
H=\left(  -i\frac{\partial}{\partial\varphi}-\hat{\mathit{\phi}}\right)
^{2}+\boldsymbol{\Omega}(\varphi)\cdot\boldsymbol{\sigma}%
.\label{HaminonAbelianContinuous}%
\end{equation}
This Hamiltonian (\ref{HaminonAbelianContinuous}) describes a
pseudo-spin moving in a one-dimension ring subject to non-Abelian
gauge field including a spin dependent flux
$\hat{\mathit{\phi}}=(\sigma_{z}+1)/4$ and an inhomogeneous magnetic
field $\boldsymbol{\Omega}(\varphi)$. Here, the natural unit is
chosen. In this sense, the induced magnetic flux $\hat{\mathit{\phi
}}$ is an operator, which does not commutate with the Zeeman term
$\boldsymbol{\Omega}\cdot\boldsymbol{\sigma}$, and the gauge field
is called non-Abelian.

\begin{figure}[ptb]
\includegraphics[bb=8 3 370 293, angle=0, width=7cm,
clip]{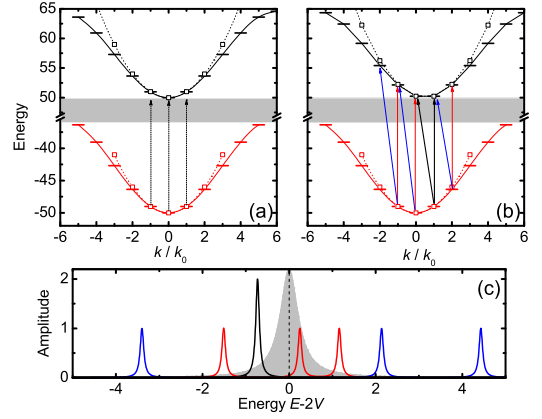}\caption{Energy levels (the solid lines
with dash marks) for the ordinary ring (a) and the M\"{o}bius ladder
(b). The dotted lines with empty square marks are the energy levels
in the continuous limit. The allowed transitions induced by a
electric field in the $z$ direction are depicted by the arrows. (c)
The optical spectra of the M\"{o}bius ladder corresponding to the
transitions shown in (b). The gray peak centered at the origin is
the spectra for the ordinary ring. The Lorentz profile are assumed
for each peaks with phenomenological broadening. Parameters used in
calculation are
$N=12$ and $V=50$.}%
\label{Fig2Levels}%
\end{figure}\textit{Topological Stark shift and spectral splitting}.- Now, we
consider the M\"{o}bius molecule subject to a uniform electric field
$\mathbf{E}=E_{z}\hat{\mathbf{e}}_{z}$ along $z$ direction. The electric field
induced on-site energy difference is $2\varepsilon_{j}=\mathbf{E}%
\cdot(\mathbf{r}_{j+}-\mathbf{r}_{j-})\equiv2\varepsilon\cos(\varphi_{j}/2)$.
By assuming the homogeneous coupling $V_{j}\equiv V$, the effective
magnetic field
\begin{equation}
\ \boldsymbol{\Omega}_{j}=\left(  \frac{\varepsilon}{2}(1+\cos\varphi
_{j}),\frac{\varepsilon}{2}\sin\varphi_{j},V\right)
\end{equation}
possesses a texture-like distribution [Fig.\ref{Fig1}(b)] with
spatially varying amplitude and direction.

The Stark effect, i.e. the energy shift under the weak electric
field, is calculated by the perturbation approach. For simplicity of
notation, we present the result in the continuous limit. By taking
the unperturbed
Hamiltonian $H_{0}=(-i\partial_{\varphi}-\hat{\phi})^{2}+\Omega_{z}%
\sigma_{z}$, the zero-th order eigen-energy are $E_{n\uparrow
}=(n-1/2)^{2}+V $ and $E_{n\downarrow}=n^{2}-V$ respectively,
corresponding to eigen-states denoted as
$|n,\chi\rangle=|n\rangle|\chi\rangle$, for
$\chi=\uparrow,\downarrow$ and $\langle\varphi|n\rangle=\exp(in\varphi)%
/\sqrt{2\pi}$. The energy spectra of these unperturbed states,
illustrated in Fig.(\ref{Fig2Levels}) in comparison with the
ordinary ring, shows the obvious spectral splitting.

The perturbation $H^{\prime}=\Omega_{x}(\varphi)\sigma_{x}+\Omega_{y}%
(\varphi)\sigma_{y}$ results in the superposition of the zero-th
order states with different pseudo-spin components. The Stark shifts
$\delta E_{n\chi}$ under the static electric field are calculated as
\begin{subequations}
\begin{align}
\delta E_{n\uparrow}  &  =\frac{\varepsilon^{2}(V-n-1/8)}{3n^{2}%
-8Vn+(4V^{2}-V-3/16)},\\
\delta E_{n\downarrow}  &  =\frac{-\varepsilon^{2}(V-n+5/8)}{3n^{2}%
-(8V+3)n+(4V^{2}+5V+9/16)}.
\end{align}
Together with the non-vanishing transition matrix elements $\langle
n\uparrow|H^{\prime}|n+1\downarrow\rangle=\langle
n\uparrow|H^{\prime }|n\downarrow\rangle=\varepsilon/2,$ this energy
shifts due to the twisted boundary condition are regarded as an
observable quantum effect of the induced non-Abelian gauge field.

In view of the future experiments, the topology state of the
M\"{o}bius molecule may be detected by the splitting of the optical
spectra. We consider the transitions of the electron in the
M\"{o}bius molecule under an optical excitation. We assume that the
molecules are subject to the linearly
polarized light, whose electric field component $\mathbf{E}(t)=E_{z}%
(t)\hat{\mathbf{e}}_{z}$ oscillates in the $z$ direction. The time
dependent Hamiltonian is $H(t)=H_{0}+\tilde{H}^{\prime}(t)$, where
the perturbation
$\tilde{H}^{\prime}(t)=\mathbf{E}(t)\cdot\mathbf{r}=H^{\prime}\cos\omega
t$, for $\omega$ being the frequency of the pumping light. The
transition selection rule shown in Fig.(\ref{Fig2Levels}) follows
from the above mentioned transition matrix elements as (i)
$|n\uparrow\rangle\rightleftharpoons|n\downarrow\rangle$ and (ii)
$|n\uparrow \rangle\rightleftharpoons|n+1\downarrow\rangle$. The
Fermi's golden rule is applied to calculate the spectra with various
excitation energy. In comparison with the case of ordinary ring,
where only one peak is located at frequency $\hbar\omega=2V$ since
the only transition (i) is allowed in this case, the optical spectra
of M\"{o}bius molecule show clear splitting due to the nontrivial
topology.

\begin{figure}[tb]
\begin{minipage}[c]{0.2\textwidth}
\centering
\includegraphics[bb=26 15 400 272, angle=0, width=5.5cm,
clip]{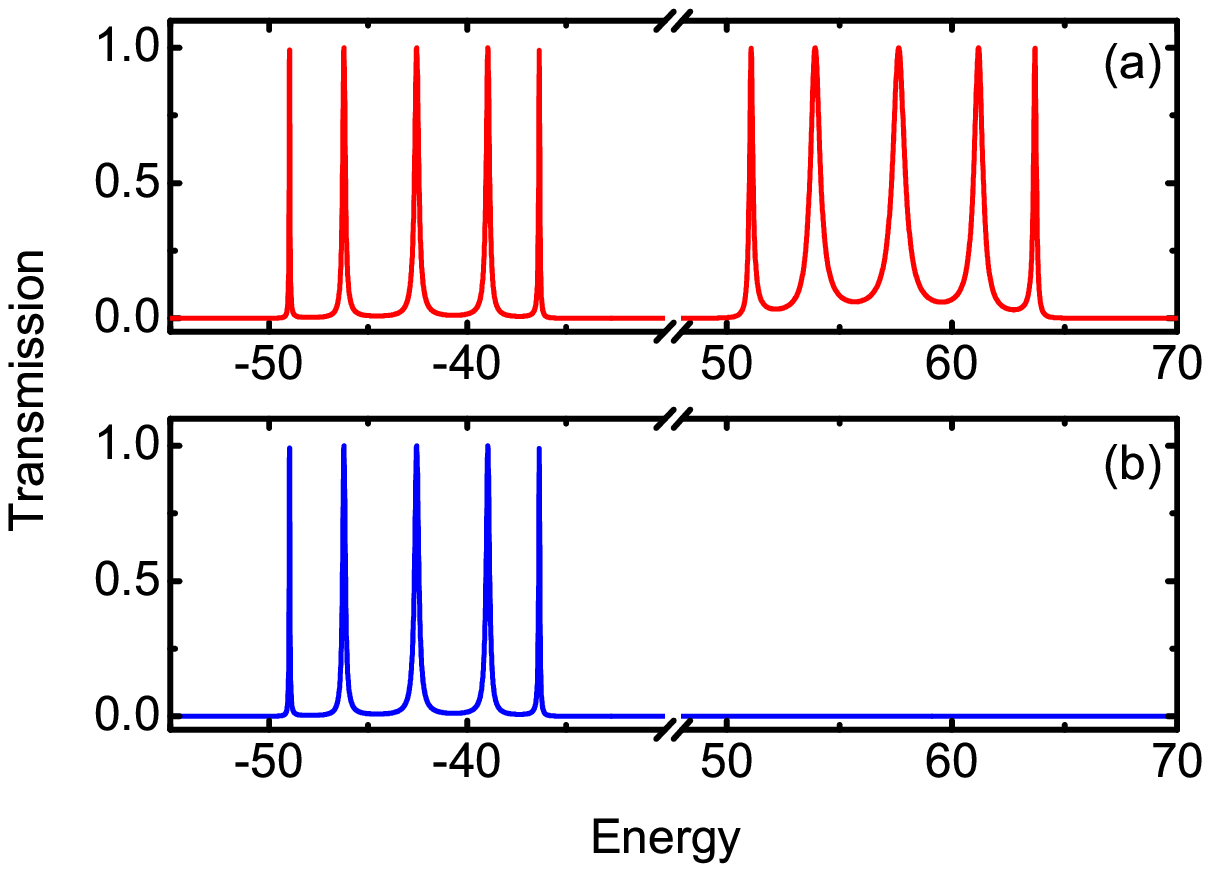}
\end{minipage}%
\begin{minipage}[c]{0.35\textwidth}
\centering
\includegraphics[bb=90 273 517 733, angle=0, width=3cm,
clip]{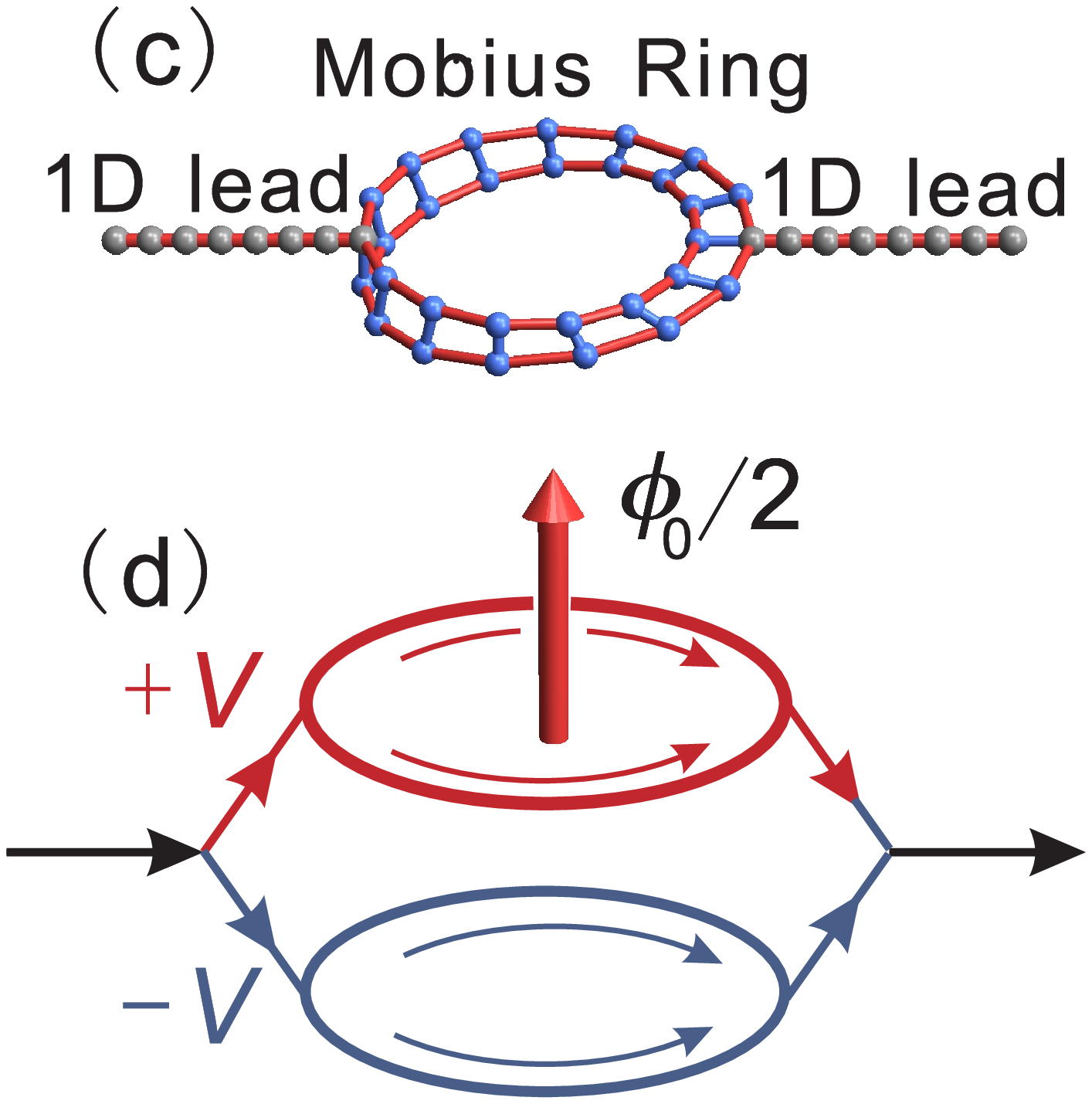}
\end{minipage}
\caption{Transmission spectra for the ordinary ring (a) and the
M\"{o}bius ring (b). (c) Schematic illustration of the M\"{o}bius
molecule joining with two leads. (d) Schematic illustration of the
topology induced magnetic flux in the ``conduction band'' (see
text). The same parameters are used as in
Fig.{\ref{Fig2Levels}}.}\label{FigTrans}
\end{figure}

\textit{Transmission through a M\"{o}bius ring}.- Besides the
topology induced Stark shift and the spectral splitting, the
transmission through the M\"{o}bius ring exhibits obvious
differences from the ordinary ring\cite{PapersOnMobius6}.

We consider the molecule is connected to two leads as shown in Fig.\ref{FigTrans}%
(c). The leads are modeled by two semi-infinite chains. The electron
can hop along the chains and tunnel between the lead and the
M\"{o}bius ring at the junctions. By assuming that the leads are
connected with the ring at $a_{0}$ and $a_{N/2}$ sites, the
Hamiltonians of the electron moving in the leads and its tunneling
to the M\"{o}bius ring the are written as
$H_{\text{lead}}=t_{l}\sum_{k=1}^{\infty
}c_{i,k}^{\dagger}c_{i,k+1}+h.c.$ and
$H_{\text{tun}}=t_{l}(c_{L,1}^{\dagger
}a_{0}+c_{R,1}^{\dagger}a_{\frac{N}{2}})+h.c.$, respectively, where
$c_{i,k}$ for $i=L,R$ are the annihilation operators of the electron
in the leads, and $t_{l}$ describes the electron hopping amplitude.

To analyze the transmission for a given injection energy $E$, we
calculate the self-energies $\Sigma_{L,R}$ to determine the Green's
function of the M\"{o}bius ring
$G(E)=[E-H-\Sigma_{L}-\Sigma_{R}]^{-1}$ by taking account of the
influence of the semi-infinite leads\cite{bookDatta}. The
self-energies $\Sigma_{L,R}$ can be obtained numerically, and they
give the level broadenings
$\Gamma_{L,R}=-2\operatorname{Im}\Sigma_{L,R}$. The transmission
coefficient $T(E)$ is obtained by the relation
$T(E)=\operatorname{Tr}[\Gamma_{R}G\Gamma_{L}G^{\dagger}]$\cite{bookDatta}.

The transmission spectrum of the M\"{o}bius ring is shown in
Fig.\ref{FigTrans}(b), comparing with that of the ordinary
mesoscopic ring with periodic boundary conditions
Fig.\ref{FigTrans}(a). For simplicity, $\varepsilon_{j}=0$ and
$V_{j}=V$ are assumed. Through the unitary transformation from
$\mathbf{A}_{j}$ to $\mathbf{B}_{j}$, the M\"{o}bius ladder is
decomposed into two independent rings (channels), see
Fig.\ref{FigTrans}(d). In the strong coupling limit, i.e. $V\gg
\xi$, the energy spectrum of the channels are split into two bands.
The energy gap between them is determined by the coupling strength
$V$. Below the energy gap (or in the ``valence band''), the
transmission behaviors are locally similar in both cases. It is not
affected by the topology of configuration space, since, as discussed
before, the induced gauge field does not present in the valence band
channels. Above the energy gap (or in the ``conduction band''), the
transmission coefficient is completely suppressed in the M\"{o}bius
ring due to the induced gauge field, which equals to a half magnetic
flux quanta. The particle could not transmit through the M\"{o}bius
ring at such energies, due to the destructive interference between
the two arms of the ring.

\textit{Decoherence from induced Stern-Gerlach effect of
pseudo-spin}.- The third topological phenomenon is the quantum
decoherence of the pseudo-spin caused by the Stern-Gerlach effect of
the induced gauge field\cite{LiYong}. Actually, through the
spin-orbit coupling, quantum entanglement between different spin
states is created so that a quantum measurement can be realized.
Similar to the Stern-Gerlach experiment, the spatial degrees of
freedom interacts with spin in a non-demolition fashion, and thus
measures the spin states. Here, we point out that, the situation may
be different from the topologically nontrivial case without obvious
local coupling. We have shown that the gauge field can be induced by
the M\"{o}bius boundary condition, and the effective pseudo-spin
orbital interaction further arises from this gauge field. Thus, the
entanglement could be created by the topology induced effect in the
absence of any \textit{real} local interactions.

In order to emphasize the main physical mechanism in our argument, we assume
the homogeneous conditions $\varepsilon_{j}=0$ and $V_{j}=V$ . Thus, only is
the $\sigma_{z}$ component retained, and thus \ Hamiltonian
(\ref{HaminonAbelian}) is of \ a block-diagonal form, i.e., $H=\text{diag}%
[H_{\uparrow},H_{\downarrow}]$ with the conditional Hamiltonians
\end{subequations}
\begin{equation}
H_{\chi}=\pm V\sum_{j=0}^{N-1}c_{j\chi}^{\dagger}c_{j\chi}%
-\sum_{j=0}^{N-1}\left(\xi_{\chi}c_{j\chi}^{\dagger}c_{j+1\chi}+h.c.\right),
\end{equation}
where $\chi\in\{\uparrow,\downarrow\}$, $\xi_{\downarrow}=\xi$, and $\xi_{\uparrow}%
=\xi\exp(i\pi/N)$.

\begin{figure}[ptb]
\includegraphics[bb=38 105 390 271, angle=0, width=7 cm,
clip]{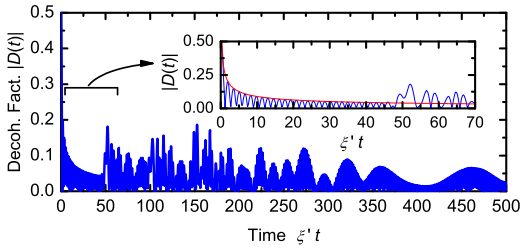}\caption{The decoherence factor as a
function of time. Inset: detailed behavior of the decoherence factor
(blue line), and its
asymptotic envelope (red line) for small $t$. $N=50$ is assumed.}%
\label{Fig5}%
\end{figure}

We assume one electron is initially located at the $a_{0}$ site,
i.e.
$|\psi(0)\rangle=a_{0}^{\dagger}|\text{vac}\rangle=|j=0\rangle\otimes\left(
|\uparrow\rangle+|\downarrow\rangle\right)  /\sqrt{2}$. In the
pseudo-spin representation, this initial state stands for a
pseudo-spin pointing in the $x$ direction at the $j=0$ site.
Obviously, it is a locally factorized state. At time $t$, the wave
function evolves into a superposition $|\psi
(t)\rangle=\left(  |\psi_{+}(t)\rangle|\uparrow\rangle+|\psi_{-}%
(t)\rangle|\downarrow\rangle\right)  /\sqrt{2}$ where the
$|\psi_{\pm }(t)\rangle=\exp\left(  -iH_{\pm}t\right)  |j=0\rangle$
can be understood as the \textit{detector states} measuring the
pseudo-spin states. As the so-called decoherence factor , the
overlap $D(t)=\langle\psi_{-}(t)|\psi
_{+}(t)\rangle=\sum_{j=0}^{N-1}G_{-}^{\ast}(j,t)G_{+}(j,t)/2$ of
$|\psi_{\pm }(t)\rangle$ characterizes the quantum coherence of
pseudo-spin states. Here,
\begin{equation}
G_{\pm}(j,t)\equiv\langle j|\psi_{\pm}(t)\rangle=\sum_{k=-\infty}^{+\infty
}e^{i\varphi_{\pm}(j_{k})}J_{j_{k}}(2\xi t),
\end{equation}
are the propagators of the two spin components, where $J_{n}(x)$ is
the $n$-th order Bessel function. Here, $j_{k}=j+kN$ represents the
$j$-th site with winding number $k$ with respect to the two
decomposed rings, and the two pseudo-spin dependent phases
$\varphi_{+}(j_{k})=j_{k}\pi/2$ and $\varphi_{-}%
(j_{k})=\varphi_{+}(j_{k})+j_{k}\pi/N$ accompany the longitudinal
motions in the two rings. It is the induced phase shift $j_{k}\pi/N$
between $\varphi_{\pm}(j_{k})$ that gives rise to the decoherence of
pseudo-spin. Straightforwardly, the decoherence factor is calculated
as
\begin{equation}
D(t)=\frac{1}{2}e^{2iVt}\sum_{\delta=-\infty}^{\infty}i^{\delta}J_{\delta
N}(2\xi^{\prime}t)
\end{equation}
where $\xi^{\prime}=\xi\sqrt{2-2\cos(\pi/N)}$. It is clear that, in
the limit with large $N$, the short time behaviors of the
decoherence factor is dominated by only one term with $\delta=0$,
thus $\left\vert D(t)\right\vert \approx\left\vert
J_{0}(2\xi^{\prime}t)\right\vert $ with an time dependent envelop
$\sqrt{N/(2\pi\xi t)}$ decaying as inverse square-root of time.

\textit{Conclusion}.- Taking the M\"{o}bius ladder as an
illustration, we have explored the role of topological structure of
the configuration space on the quantum states of the particles
confined in a topologically-nontrivial manifold. The global
properties of the topological system can be locally described by a
non-Abelian gauge structure, which can result in some observable
effects in the aspects of spectroscopy, such as the topological
induced Zeeman splits and the higher energy band suppression of the
transmission of the M\"{o}bius molecule. We also show the quantum
decoherence of the pseudo-spin and the entanglement due to the
pseudo-spin orbital interaction. On the other hand, from the view of
chemistry, these observable effects can be regarded as the physical
signals of the successful synthesis of some M\"{o}bius aromaticity
molecule. These methods may be used to distinguish the topologically
nontrivial molecules from the ordinary ones.

This work is supported by NSFC No.10474104, No. 60433050, and No.
10704023, NFRPCNo. 2006CB921205 and 2005CB724508. One (CPS) of the
authors acknowledges Y.S. Wu for his stimulating discussions.

\end{document}